\documentclass[journal=apchd5,manuscript=letter]{achemso}

\usepackage[version=3]{mhchem} % Formula subscripts using \ce{}
\usepackage[T1]{fontenc}       % Use modern font encodings
\usepackage{color}
\graphicspath{{figs/}}
\usepackage[export]{adjustbox}
\usepackage{natbib}
\usepackage{cuted}

\newcommand{\TM}{^{^{\scriptscriptstyle (\rm TM)}}}
\newcommand{\TE}{^{^{\scriptscriptstyle (\rm TE)}}}
\newcommand{\p}{^{^{\scriptscriptstyle (\rm \alpha)}}}

\author{Sergejs~Boroviks}
\affiliation [University of Southern Denmark]
{Center for Nano Optics, University of Southern Denmark, Campusvej 55, DK-5230~Odense~M, Denmark}
\email{sebo@mci.sdu.dk}

\author{Rucha~A.~Deshpande}
\affiliation [University of Southern Denmark]
{Center for Nano Optics, University of Southern Denmark, Campusvej 55, DK-5230~Odense~M, Denmark}
\author{N.~Asger~Mortensen}
\affiliation [University of Southern Denmark]
{Center for Nano Optics, University of Southern Denmark, Campusvej 55, DK-5230~Odense~M, Denmark}
\alsoaffiliation[]{Danish Institute for Advanced Study, University of Southern Denmark, Campusvej 55, DK-5230~Odense~M, Denmark}
\author{Sergey~I.~Bozhevolnyi}
\affiliation [University of Southern Denmark]
{Center for Nano Optics, University of Southern Denmark, Campusvej 55, DK-5230~Odense~M, Denmark}
\alsoaffiliation[]{Danish Institute for Advanced Study, University of Southern Denmark, Campusvej 55, DK-5230~Odense~M, Denmark}
\email{seib@mci.sdu.dk}

\title[PSFMM]{Multifunctional meta-mirror: polarization splitting and focusing}

\abbreviations{GSP, MIM, PSFMM, EBL, SEM, CCD, PER}

\begin{document}

\begin{abstract}
Metasurfaces are paving the way to improve traditional optical components by integrating multiple functionalities into one optically flat metasurface design. We demonstrate the implementation of a multifunctional gap surface plasmon-based metasurface which, in reflection mode, splits orthogonal linear light polarizations and focuses into different focal spots. The fabricated configuration consists of 50\,nm thick gold nanobricks with different lateral dimensions, organized in an array of 240\,nm$\times$240\,nm unit cells on the top of a 50\,nm thick silicon dioxide layer, which is deposited on an optically thick reflecting gold substrate. Our device features high efficiency (up to $\sim 65$\%) and polarization extinction ratio (up to $\sim 30$\,dB),  exhibiting broadband response in the near-infrared band (750--950\,nm wavelength) with the focal length dependent on the wavelength of incident light. The proposed optical component can be forthrightly integrated into photonic circuits or fiber optic devices.
\end{abstract}

Optical metasurfaces\cite{kildishev2013a,Yu:2014a,zhao2014a,yu2015a} -- planar (quasi-two-dimensional) sub-wavelength artificial metallic and dielectric structures -- have developed tremendously in recent years\cite{Sci:Yu:2011,rogers2012a,Arbabi:2015,High2015,Lin2014,Chong2015,Zhu:2017a}. With the progress in nano-fabrication methods, different applications of metasurfaces were demonstrated experimentally, ranging from artificial plasmonic coloring\cite{Kristensen:2016} to flat optical components \cite{NanoLetters:Pors:2013,OptExp:Pors:2013_Qurterwp,SciRep:Pors:2013,Ding:2017b,Wu:2017}. In fact, many of demonstrated functionalities cannot be realized with conventional (bulk) optical components (see recent reviews\cite{Ding:2017,PhysRep:Glybovski:2016,Hsiao:2017}). 
An important aspect of the development of flat optics is efficient integration of diverse functionalities into a single component of a sub-wavelength thickness, which is also not attainable with conventional diffraction limited optical components.\cite{Sci:Yu:2011} Thus, the scope of current research was recently drawn to exploration of \emph{multifunctional} metasurfaces \cite{Maguid:2016, Cheng:2017,Maguid:2017,Forouzmand:2017}, including reconfigurable designs. \cite{Wu:2017:AOP}

\begin{figure}[htb!]
	\begin{center}
		\includegraphics[width= 0.65\columnwidth]{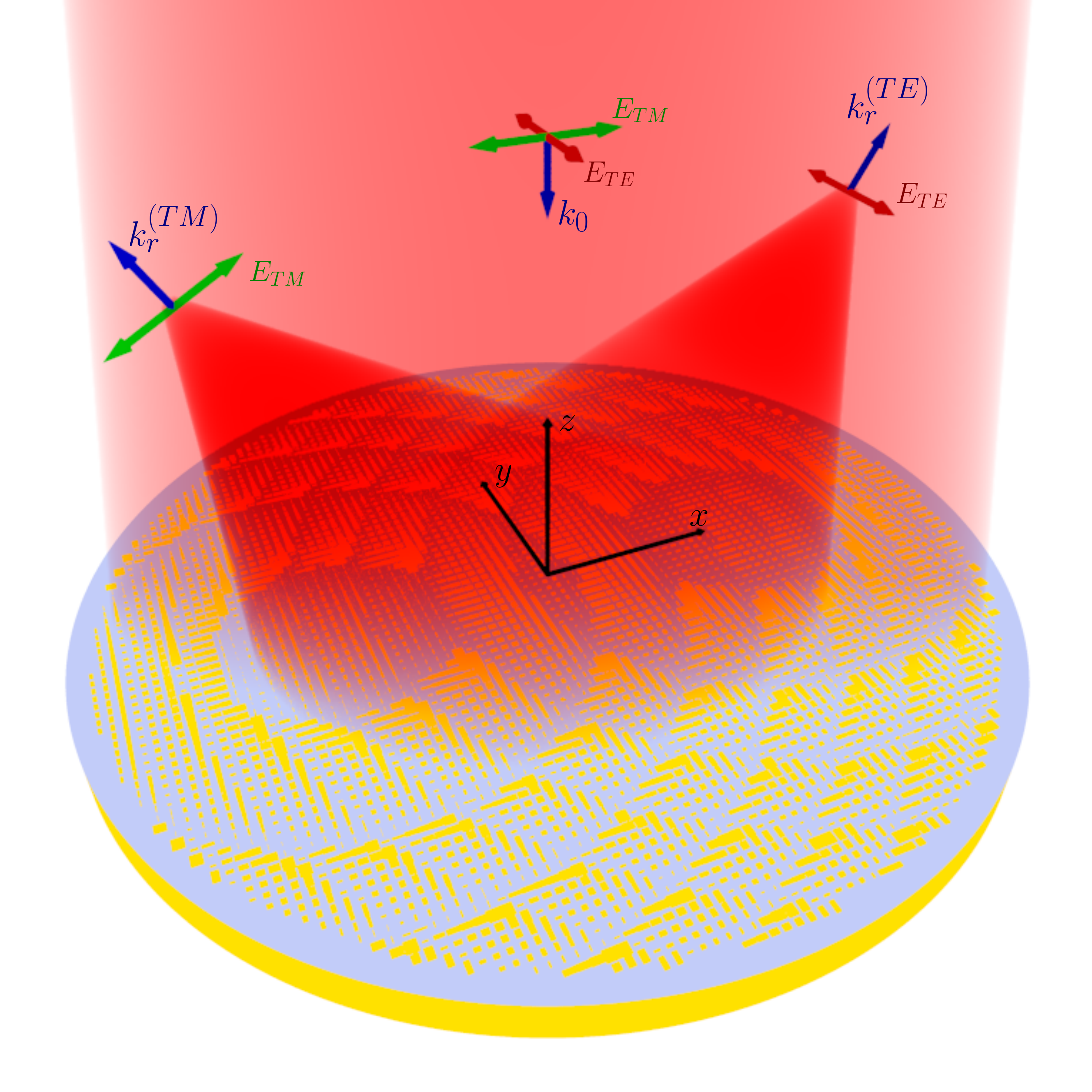}
		\caption{Illustration of the working principle (artistic rendering) of a multifunctional metasurface device. When illuminated by linearly polarized light, the device focuses and splits orthogonal linear light polarizations into different focal spots with high efficiency and a high polarization extinction ratio.} 
		\label{fig:MSWP}
	\end{center}
\end{figure}

For polarization-controlled optical systems, for example polarization multiplexed fiber-optic communications \cite{Core:2006} or polarization-assisted sensing\cite{Caucheteur:2017}, single-chip multi-functionality is highly advantageous. Here, we demonstrate design of a multifunctional metasurface, which can be straightforwardly used in such systems. It functions as a polarization-sensitive parabolic reflector (hereinafter referred to as PSFMM -- Polarization Splitting and Focusing Meta-Mirror), simultaneously splitting orthogonal light polarizations and focusing into different focal spots at the design wavelength $\lambda=800$\,nm (illustration of the working principle is shown in Figure~\ref{fig:MSWP}). Previous designs targeted splitting of orthogonal polarizations in reflection and transmission and were implemented using dielectric or semiconductor metasurfaces\cite{Arbabi:2015,RSC:Guo:2017,APL:Lee:2014}, or alternatively in radio-frequency band\cite{ADOM:Cai:2017,APA:Guo:2016,Xu:2016}. 

Our device utilizes gap surface plasmon (GSP) resonators\cite{Miyazaki:2006,Kurokawa:2007,Bozhevolnyi:2007,Sondergaard.2008a} as constitutive elements, which support highly localized plasmonic resonances that form flexible meta-atom building blocks of metasurfaces,\cite{kildishev2013a,Yu:2014a,Roberts:2014,Ding:2017,PhysRep:Glybovski:2016,Hsiao:2017} with the possibility to engineer the local phase and reflection amplitude in gradient metasurfaces.\cite{OptExp:Pors:2013,Ding:2017} GSP-based metasurfaces operate in the reflection mode, which enables high efficiency,\cite{SciRep:Pors:2013} reaching $\sim 80\%$ for various applications \cite{Sun:2012,Chen:2014,Zheng:2015}. 

\begin{figure}[htb!]
\begin{center}
\includegraphics[width= 1\columnwidth]{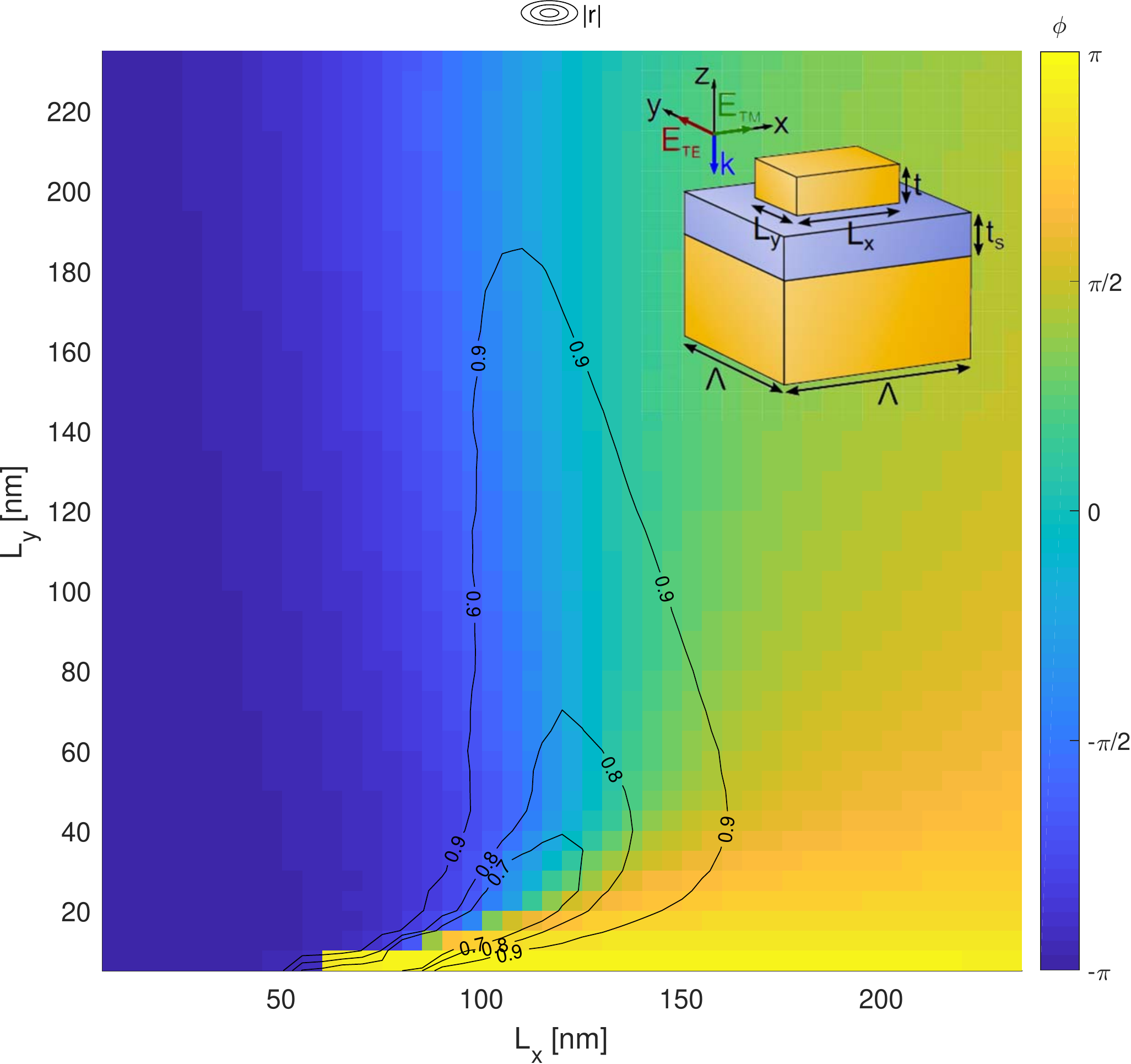}
\caption{Simulated reflection coefficient $r=|r|\exp(i\phi)$, with the phase $\phi=\arg (r)$ shown in colour map as a function of nanobrick lateral dimensions $L_x$ and $L_y$ for $\Lambda= 240$\,nm, $t=t_s=50$\,nm for TM polarized incident light at $\lambda=800$\,nm wavelength. Contour lines indicate the corresponding reflection coefficient amplitude $|r|$. Reflection phase and amplitude corresponding to TE polarization are obtained by transposing this map. (inset) Sketch of the metasurface unit cell with indicated dimensions.} 
\label{fig:GSPmap}
\end{center}
\end{figure}

The basic element of the metasurface, also referred to as the unit cell, has a period of $\Lambda=240$\,nm and is comprised of lithographically patterned gold nanobricks of height $t=50$\,nm and lateral dimensions $L_x$ and $L_y$. The nanobricks are supported by a silicon dioxide layer ($t_s=50$\,nm) deposited on optically thick gold substrate, schematically depicted in the inset of Figure~\ref{fig:GSPmap}. Such a metal-insulator-metal (MIM) configuration is, in its nature, a GSP resonator which is known to exhibit strong field confinement in the dielectric layer under the metal nanobrick. Due to this property, a negligible coupling between neighboring unit cells is permissibly assumed,\cite{SciRep:Pors:2013} which facilitates the construction of phase gradient metasurfaces.\cite{Ding:2017}

The electrodynamics of the considered MIM configuration is modeled using a finite-element method (Comsol Multiphysics), assuming that perpendicularly incident light (of wavelength $\lambda=800$\,nm), is polarized along the $x$-axis. Parametric sweeps are run through all possible combinations of nanobrick lateral dimensions $L_x$ and $L_y$ from 5 to 235\,nm, in steps of 5\,nm, calculating complex reflection coefficient $r=|r|\exp(i\phi)$. The length of the step was chosen to comply with the resolution capability of the electron-beam lithography (EBL) used in our later fabrication of samples.
Two degrees of freedom in the design geometry, i.e. the lateral dimensions of the gold nanobrick, give control over reflection phase, $\phi$, in almost full $2\pi$ phase space near the GSP resonance, as shown in colourmap in Figure~\ref{fig:GSPmap}. Naturally, it is possible to control phase response of the unit cell independently for two orthogonal linear light polarizations -- transverse magnetic (TM) or $x$-polarization, shown explicitly in the Figure~\ref{fig:GSPmap}, and transverse electric (TE) or $y$-polarization, being equivalent to the transpose of the map shown in Figure~\ref{fig:GSPmap}. 
Except for a narrow region of dimensions close to the resonance, our generic GSP resonator facilitates a high reflection amplitude, $|r|$, (see contour map in Figure~\ref{fig:GSPmap}), which is a crucial property for the overall meta-mirror efficiency. 

\section{Design considerations}

In general, metasurfaces can be designed to mimic compact Fresnel reflectors by imposing a hyperboloidal phase profile on them.\cite{NanoLetters:Aieta:2012} The underpinning of this idea can be shown by investigating the optical path difference of rays reflected from the metasurface: as incident light is assumed to be a plane wave with a harmonic time dependence, $\mathbf{E} \left( \mathbf{x},t \right)=\mathbf{E_0}\exp[i(\mathbf{k}\cdot \mathbf{x}-\omega t)]$, the reflected rays differ from it only by a factor $\exp[i\phi_{\scriptscriptstyle(\alpha)}]$, which is the reflection phase ($\alpha$ indicates the polarization, i.e. either TM or TE). Phase accumulated in the redundant optical path can be compensated by letting the reflection phase on the metasurface follow a hyperboloidal profile, proportional to the incident wave number. 

\begin{figure*}[htb!]
	\includegraphics[width= 1\textwidth]{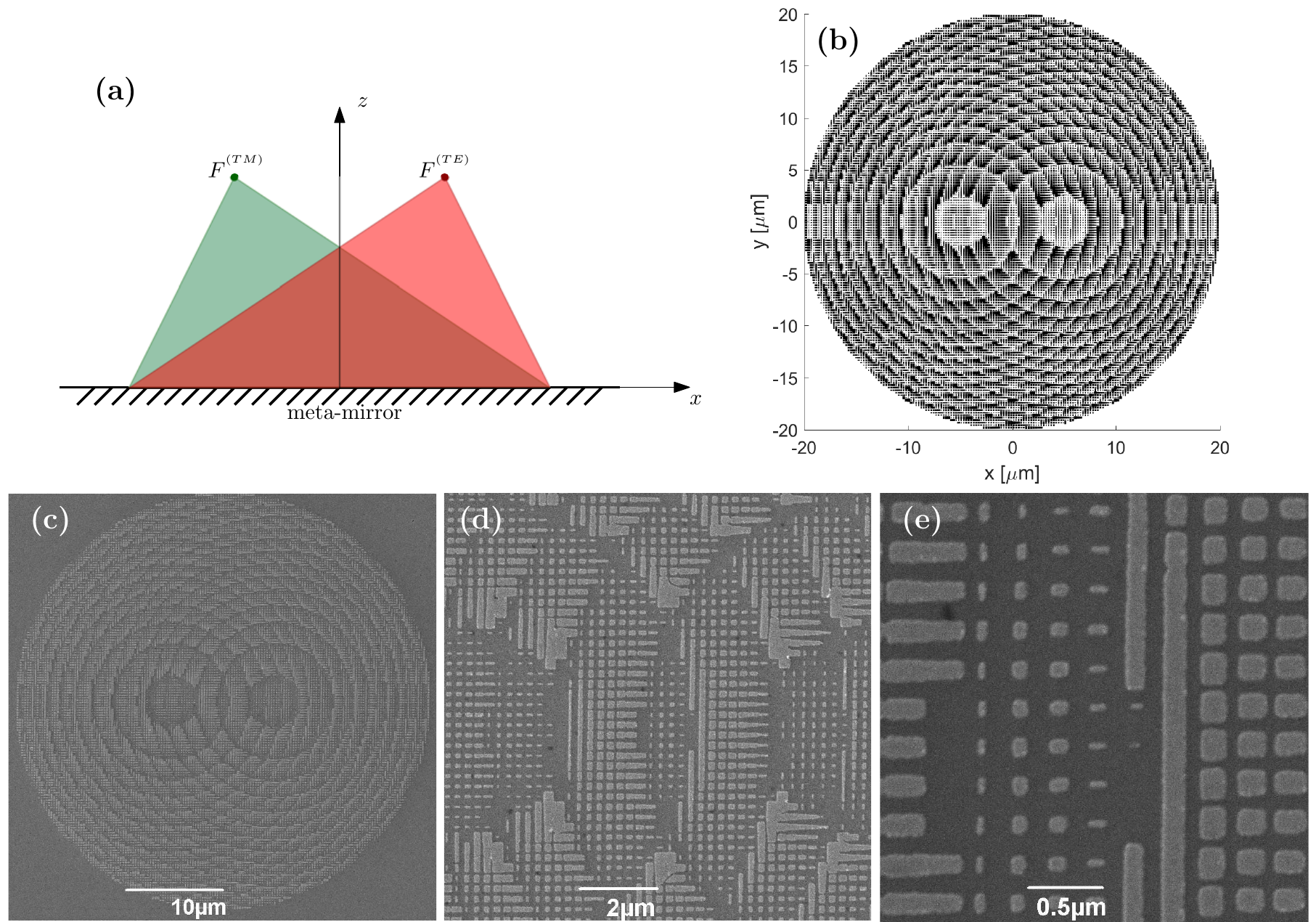}
		\caption{Design of the polarization splitting and focusing meta-mirror: (a) Locations of focal points $F\TM= (d_x\TM,d_y\TM,d_{z}\TM)$ and $F\TE=(d_x\TE,d_y\TE,d_{z}\TE)$  for orthogonal linear light polarizations; (b) geometry of the PSFMM design; (c)-(e) SEM images of the fabricated PSFMM sample at  different magnifications with a zoom-in on the single meta-atom building blocks of the metasurface.} 
		\label{fig:PSFMMdes}
\end{figure*}

In our case, for simultaneous polarization splitting and focusing, two phase profiles are needed, one for each polarization, with centres of hyperboloids shifted one from another to achieve off-axis focusing. Thus, the reflection phase $\phi=\arg (r)$ in every cell of the metasurface must simultaneously satisfy two conditions:
\begin{equation}
\phi_{\scriptscriptstyle(\alpha)} (x,y)=\frac{2\pi}{\lambda} \left(d_{z}\p-\sqrt{\left(x-d_{x}\p\right)^2+\left(y-d_{y}\p\right)^2+\left(d_{z}\p\right)^2}\right)
\end{equation}
where $d_{z}\TM$ and $d_{z}\TE$, $d_{x}\TM$ and $d_{x}\TE$, $d_{y}\TM$ and $d_{y}\TE$ are coordinates of the focal points for TM and TE polarizations along $z$, $x$ and $y$-axis respectively, as depicted in Figure~\ref{fig:PSFMMdes}a. Similar approach for achieving off-axis focusing was also used in different contexts \cite{Deng:16,Chen:2017}.

In contrast to previous works,\cite{NanoLetters:Pors:2013,SciRep:Pors:2013,NanoLetters:Aieta:2012} we do not limit the choice of metasurface constitutive elements to some relatively small discrete design space. Instead, appropriate $L_x$ and $L_y$ parameters for each unit cell are chosen from the entire space of simulated values (Figure~\ref{fig:GSPmap}), namely from the array of size $47\times 47$ elements. Even though it makes the optimization of our lithographic fabrication more complicated, the phase gradient obtained with this approach is closer to ideal, i.e. deviations from the imposed hyperboloidal phase profile introduced by fabrication imperfections are less pronounced compared to the case when the deviation is introduced also by a relatively significant phase steps due to limited number of constitutive elements, although elements themselves are perhaps fabricated with better tolerances. 
One could also increase the variety of $L_x$ and $L_y$ parameters  (e.g. by interpolation within simulated values of $L_x$ and $L_y$), to make mapping to the ideal phase profile even more accurate. However, this would be an unnecessary complication, since the discretization of the possible element dimensions is practically limited by the resolution of EBL equipment ($\sim 5$\,nm). Also, we do not impose any limitation on minimal reflection amplitude of the constitutive elements, as elements with low $|r|$ for TM polarization have high value of the same parameter for TE (and vice versa), which on average results in reasonably good overall reflectivity and better correspondence to the imposed phase gradient. 

The geometry of the PSFMM design selected for experimental investigation, with focusing parameters set to $d_{z}\TM=d_{z}\TE=15$\,$\rm \mu$m, $d_{x}\TM=-d_{x}\TE=5$\,$\rm \mu$m, and $d_{y}\TM=d_{y}\TE=0$, is shown in Figure~\ref{fig:PSFMMdes}b. This design diverges orthogonal polarizations by an angle of $\sim 37^\circ$ and focuses them at a distance of 15\,$\rm\mu$m from the metasurface. The metasurface design region is circular with diameter $D=40$\,$\rm\mu$m. The diameter defines the focusing ability of the meta-mirror, quantitatively measured in terms of the numerical aperture $\text{NA}\approx\sin\left[\tan^{-1}(D/2d_{z}\p)\right]$ (evaluated to be 0.8 for this design at the design wavelength $\lambda=800$\,nm).

\section{Experiment}

The multifunctional meta-mirror sample is fabricated using standard EBL and lift-off techniques. First, the substrate is prepared: a 150\,nm-thick layer of Au and 50\,nm of SiO$_2$, with 3\,nm-thin titanium layers in between for adhesion purpose, is deposited on a Si wafer, using thermal evaporation (Cryofox TORNADO 405 evaporation system by Polyteknik) for gold and titanium layers, while RF-sputtering is employed for SiO$_2$. Furthermore, a 200\,nm layer of PMMA 950 A2 resist (MicroChem) is spin-coated, which is used as a stencil material for creating nanobrick structures. The PSFMM design pattern is then created using EBL (JEOL-640LV SEM with an ELPHY Quantum lithography attachment). 
Gold nanobricks are formed by thermal evaporation of 3\,nm of Ti and 50\,nm of Au followed by lift-off process (etching away stencil material and development). As can be seen from scanning electron microscope (SEM) images in Figure~\ref{fig:PSFMMdes}c-e, apart from the smallest features, fabrication quality of the resulting sample is in overall accordance with our initial design requirements also at the level of the meta-atom building blocks of the metasurface. 

\begin{figure}[htb!]
	\includegraphics[width= 0.6\columnwidth]{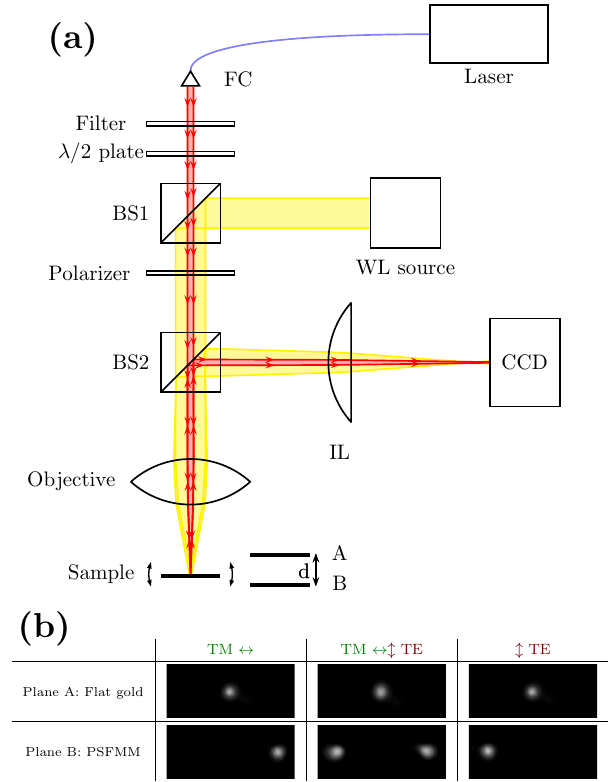}
		\caption{(a) Schematic diagram of the experimental setup for optical characterization of the fabricated sample; (b) images of spots (brightness adjusted for better visibility) produced by flat gold surface and PSFMM correspondingly in the planes A and B, when illuminated by laser beam at $\lambda=800$\,nm wavelength with different polarizations.} 
		\label{fig:expstp}
\end{figure}

Optical characterization of the sample is performed using a tunable Ti-Sapphire laser (3900S CW by Spectra-Physics), whose light is directed through a neutral density (ND) filter, a combination of a Glan--Thompson and a half-wave plate and two beam splitters to an $\times60$\,objective (Edmund Optics, NA=0.85, Achromatic, 0.15\,mm working distance, chosen to have higher NA than the designed metasurface), which is used to focus light onto the sample. The reflected light is collected using the same objective and directed via a beam splitter to an imaging lens, which focuses light onto a CCD camera (Mightex CCEB013-U, monochrome), as schematically depicted in Figure~\ref{fig:expstp}a. Besides, additional white-light illumination is used for convenience of visually locating the meta-mirror on the surface of the sample.
Since the fabricated multifunctional meta-mirror is designed to exhibit a very short focal distance ($\sim15\,\mu$m), it is not practical to introduce a beam splitter between the sample surface and the objective to measure focusing characteristics with an unfocused laser beam.

However, in this setup, the focusing effect can be verified by investigating how the fabricated meta-mirror and flat unstructured gold surface would reflect light when being placed at different distances from the objective. From geometrical optics, it can be shown that the plane (referred to as plane A in Figure~\ref{fig:expstp}) at which the flat gold surface produces a focused spot on the CCD camera screen, is located at a distance $d=2d_{z}\p$ away from the plane B, at which the PSFMM results in a focused (deflected) spots on the screen. 

Optical images of the spots created by the flat gold surface and multifunctional meta-mirror, when illuminating the sample with $\lambda=800$\,nm light at different polarization states are shown in Figure~\ref{fig:expstp}b. As can be seen, PSFMM deflects TM polarization to the right hand-side and TE polarization to the left hand-side from the origin. Besides, raw images of the focal spots captured at different wavelengths were used to estimate the efficiency of the fabricated meta-mirror: the reflected power was determined by integrating intensity of pixels in the captured images.

\section{Results and Discussion}

\begin{figure*}[htb!]
	\includegraphics[width=0.6\textwidth]{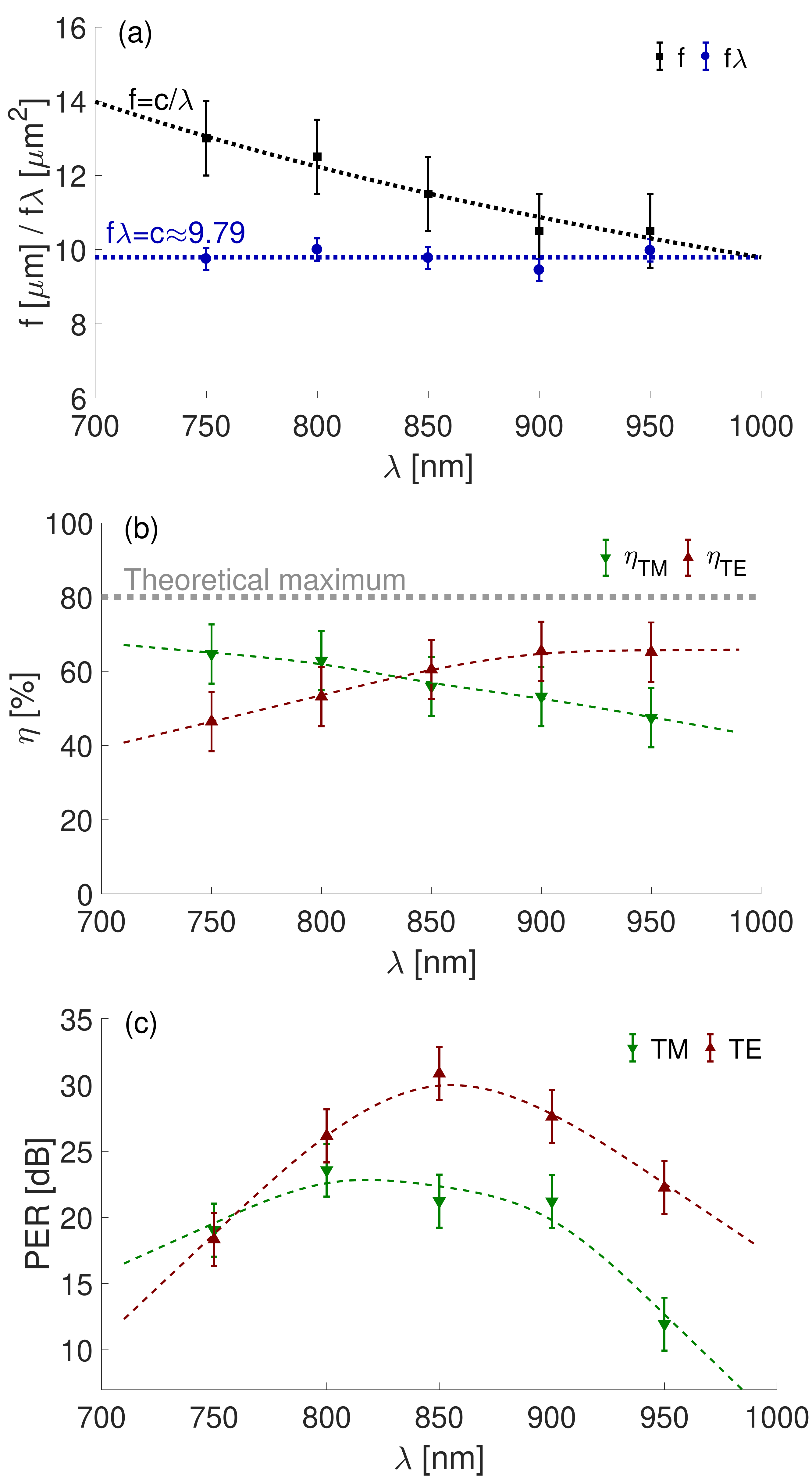}
	\caption{Performance of the fabricated PSFMM sample evaluated at five wavelengths in the 750--950\,nm range: (a) focal length $f$ and $f\lambda$ product, dotted lines emphasize correspondence with idealized trends; (b) efficiencies measured for orthogonal polarizations, idealized theoretical maximum \cite{NanoLetters:Pors:2013} indicated in grey dashed line; (c) polarization extinction ratio, defined as ratio of intensities of observed spots.} 
	\label{fig:resultss}
\end{figure*}

The fabricated polarization-splitting and focusing meta-mirror sample was characterized, demonstrating comparatively good focusing characteristics at different wavelengths in the range of 750--950\,nm, despite the inherent, but minor imperfections in fabrication (Figure~\ref{fig:PSFMMdes}c-e). Measured focal length $f$ (Figure~\ref{fig:resultss}a) is slightly smaller than its initially designed value ($\sim13$\,$\mu$m instead of 15\,$\mu$m at $\lambda=800$\,nm), however, this deviation can be explained by the spherical aberration of the objective and imaging lens (we made no special effort to account for this). As was expected from the previous theoretical work\cite{NanoLetters:Pors:2013}, the focal length of the meta-mirror decreases when increasing the wavelength. It is worth noting that such chromatic aberration effect also decreases the NA value of a meta-mirror for longer wavelength, as was also shown in a recent work on achromatic metasurfaces \cite{Wang:2017}. In turn, its $f\lambda$ product (a parameter which determines the dimensions of Fresnel zones for conventional Fresnel lenses and mirrors) is practically constant ($\sim 10\,\mu$m). This can be anticipated, since there is no strong dispersion of the constitutive materials in the considered near-infrared wavelength range. 

The efficiency, defined as the ratio of powers of the light reflected by the meta-mirror located in plane B and by a flat unstructured gold surface in plane A, reaches $\sim65\%$ in the best case, which is significantly larger than in the previous demonstrations.\cite{NanoLetters:Pors:2013,NanoLetters:Aieta:2012} Previous numerical study of 1D GSP-based focusing meta-mirrors\cite{NanoLetters:Pors:2013} predicts the ultimate efficiency to be on the level of $\sim\,80\%$, promising also the broadband operation that we observe experimentally, with modest variations of properties across our wavelength regime of interest. Numerical simulations\cite{NanoLetters:Pors:2013} do not suggest any clear trends in wavelength dependences of key observables for relatively small wavelength variations. We thus attribute the observed spectral variations in our experiments mainly to unintended fabrication variations. In fact, taking the error bars into account, most of our observables are to a first approximation wavelength independent, demonstrating high efficiency ($>45\%$) over the entire measurement range.

Finally, the polarization-extinction ratio (PER) was studied (Figure \ref{fig:resultss}c). Here, PER is defined as the ratio of intensities in \emph{correct} and \emph{incorrect} spots when illuminated with a light of indicated polarization (shown on a dB scale). As can be seen, PER is very high, reaching $~30$\,dB value, which implies that only $\sim1/1000$ (conservative estimate) of incident power is deflected into the \emph{incorrect} focal spot.
Error bars in the Figure~\ref{fig:resultss} indicate measurement uncertainty due to inaccuracy in the axial positioning of the meta-mirror during the focal length measurement  ($\pm1\,\mu$m) and unknown sensitivity of CCD camera at different wavelengths (taken to be $\pm10\,\%$, as a conservative estimate).

To summarize, we have demonstrated a broadband and efficient multifunctional metasurface, which is capable of simultaneous focusing and polarization splitting. Functionalities capitalize from highly in-plane localized GSP modes hosted in an MIM configuration, which facilitate nearly independent reflection phase gradients for two orthogonal polarizations. Our class of devices can be directly integrated into various systems employing polarization multiplexing. While we have emphasized the near-infrared regime, the design can be readjusted to operate equally efficiently also in the telecommunication wavelength window. Finally, our EBL-based demonstration can be extended for large-scale fabrication, for example using roll-to-roll printing approaches now reaching out for metasurface mass-production.\cite{Hojlund-Nielsen:2016,Murthy:2017}

\begin{acknowledgement}
S.~I.~B. acknowledges the European Research Council, Grant 341054 (PLAQNAP). N.~A.~M. is a VILLUM Investigator supported by VILLUM FONDEN (grant No. 16498). Center for Nano Optics is financially supported by the University of Southern Denmark (SDU 2020 funding). We acknowledge A.~S.~Roberts for help with fabrication issues, V.~Zenin for suggestions on optical characterization and J.~Linnet for useful comments on an early version of the manuscript.
\end{acknowledgement}

%\bibliography{metasurface}

\providecommand{\latin}[1]{#1}
\makeatletter
\providecommand{\doi}
  {\begingroup\let\do\@makeother\dospecials
  \catcode`\{=1 \catcode`\}=2\doi@aux}
\providecommand{\doi@aux}[1]{\endgroup\texttt{#1}}
\makeatother
\providecommand*\mcitethebibliography{\thebibliography}
\csname @ifundefined\endcsname{endmcitethebibliography}
  {\let\endmcitethebibliography\endthebibliography}{}

\end{document}